\documentclass[prl,twocolumn,showpacs]{revtex4}

\usepackage{amsmath}
\usepackage{amsfonts}

\newcommand{\R}{{\mathbb R}}               

\renewcommand{\Re}{\mbox{\rm Re}}

\newcommand{\w}{\omega}
\newcommand{\f}{\frac}

\renewcommand{\tilde}{\widetilde}

\begin{document}







\title{Hamiltonian approach to the dynamical Casimir effect
\footnote{Dedicated to Steve Fulling.}}

\author{Jaume Haro$^{1,}$\footnote{E-mail: jaime.haro@upc.edu} and
Emilio Elizalde$^{2,}$\footnote{Presently on leave at
Department of Physics \& Astronomy, Dartmouth College, 6127 Wilder
Laboratory, Hanover, NH 03755, USA. E-mail: elizalde@ieec.fcr.es,
elizalde@math.mit.edu}}

\affiliation{$^1$Departament de Matem\`atica Aplicada I, Universitat
Polit\`ecnica de Catalunya, Diagonal 647, 08028 Barcelona, Spain \\
$^2$Instituto de Ciencias del Espacio (CSIC) \& Institut
d'Estudis Espacials de Catalunya (IEEC/CSIC)\\ Campus UAB, Facultat
de Ci\`encies, Torre C5-Parell-2a planta, 08193 Bellaterra
(Barcelona) Spain}

\begin{abstract} A Hamiltonian approach is introduced in order to
address some severe problems associated with the physical
description of the dynamical Casimir effect at all times. For
simplicity, the case of a neutral scalar field in a one-dimensional
cavity with partially transmitting mirrors (an essential proviso) is
considered, but the method can be extended to fields of any kind and
higher dimensions. The motional force calculated in our approach
contains a reactive term ---proportional to the mirrors'
acceleration--- which is fundamental in order to obtain
(quasi)particles with a positive energy all the time during the
movement of the mirrors
---while always satisfying the energy conservation law. Comparisons
with other approaches and a careful analysis of the interrelations
among the different results previously obtained in the literature
are carried out.

\end{abstract}


\maketitle

{\it Introduction.}---Moving mirrors modify the structure of the
quantum vacuum, what manifests in the creation and annihilation of
particles. Once the mirrors return to rest, a number of the produced
particles may still remain which can be interpreted as radiated
particles. This flux has been calculated in several situations by
using different methods, as averaging over fast oscillations
\cite{dk96,jjps97}, by multiple scale analysis \cite{cdm02}, with
the rotating wave approximation \cite{sps02}, with numerical
techniques \cite{ru05}, and others \cite{bmoo}. Here we will be
interested in the production of the particles and their possible
energy all the time while the mirrors are moving. In the case of a
single, perfectly reflecting mirror, the number of produced
particles as well as their energy diverge while the mirror moves.
Several renormalization prescriptions have been used in order to
obtain a well-defined energy, however, for some trajectories this
finite energy is not a positive quantity and cannot be identified
with the energy of the produced particles (see e.g. \cite{fd76}).

Our approach relies on two basic ingredients: proper use of a
Hamiltonian method and the consideration of partially transmitting
mirrors, which become transparent to very high frequencies. We shall
prove, in this way, both that the number of created particles is
finite and also that their energy is always positive for the whole
trajectory corresponding to the mirrors' displacement. We will also
calculate the radiation-reaction force that acts on the mirrors
owing to the emission and absorption of particles, and which is
related with the field's energy through the energy conservation law,
so that the energy of the field at any time $t$ is equal, with
opposite sign, to the work performed by the reaction force up to
time $t$ \cite{fv82,be94}. Such force is usually split into two
parts \cite{e93,bc95}: a dissipative force whose
work equals minus the energy of the particles that remain
\cite{fv82}, and a reactive force vanishing when the mirrors
return to rest. We will show that the
radiation-reaction force calculated from the Hamiltonian approach
for partially transmitting mirrors satisfies, at all time, the energy
conservation law and can naturally account for the creation of
positive energy particles. Also, the dissipative part we
will obtain agrees with the one calculated by other methods, as
using the Heisenberg picture or other effective Hamiltonians. Note
that those methods have problems with the reactive part,
which in general yields a
non-positive energy that cannot be considered as that
 of the particles created at any $t$.

In what follows, we first introduce the Hamiltonian method for a
neutral Klein-Gordon field in a cavity with boundaries moving at a
certain speed $v << c$. Then, a single partially transmitting
mirror in 1+1 spacetime will be studied in order to illustrate the
procedure and prove the above statements. Our results will be
compared with the ones in the literature. Finally, the
case of two mirrors will be investigated, to see that also here we
obtain physically meaningful quantities while the mirrors move, in
an unambiguous way, and that the dissipative force does agree with
previous results by other authors. \medskip

{\it The Hamiltonian formulation.}---Consider a neutral massless
scalar field in a cavity $\Omega_t$, and assume that the boundary is at
rest for time $t\leq 0$ and returns to its initial position at time
$T$. Suppose also its velocity to be of order $\epsilon =v/c$ (dimensionless,
it is of order $10^{-8}$ in \cite{kbo1}, see later).
The Lagrangian density of the field is
${\mathcal L}(t,{\bf x})=\f{1}{2}[(\partial_t\phi)^2 -|\nabla_{\bf
x}\phi|^2], \, {\bf x}\in\Omega_t\subset \R ^3, \, t\in \R.$ In
terms of the canonical conjugated momentum $ {\xi}(t,{\bf
x})\equiv\f{\partial {\mathcal L}} {\partial( {\partial_t\phi})}=
{\partial_t\phi}(t,{\bf x}), $ the energy density of the field is
${{\mathcal E}}(t,{\bf x})\equiv {\xi}{\partial_t\phi}-{\mathcal
L}(t,{\bf x}) =\f{1}{2} \left({\xi}^2 +|\nabla_{\bf
x}\phi|^2\right),$ while its energy is
$E(t;\epsilon)\equiv\int_{\Omega_t}d^3{\bf x} \, {{\mathcal
E}}(t,{\bf x})$. As is well known, this energy density does not
coincide with the Hamiltonian one \cite{rt85}-\cite{sps98}. The
Hamiltonian density can be conveniently obtained using the method in
\cite{js96}.

First, a (non-conformal) coordinate change is used to convert the
moving boundary $\Omega_t$ into a fixed one $\tilde\Omega$:
 $(t(\bar{t},{\bf y}), {\bf x}(\bar{t},{\bf y}))$
$={\mathcal R}(\bar{t},{\bf y})=(\bar{t}, {\bf R}(\bar{t},{\bf y}))$
($\bar{t}$ the new time). The action of the system is
$S=\int_{\R}\int_{\tilde{\Omega}} d^3{\bf y}d\bar{t}\tilde{\mathcal
L}(\bar{t},{\bf y}),$ with $\tilde{\mathcal L}(\bar{t},{\bf y})
\equiv J{\mathcal L}({\mathcal R}(\bar{t},{\bf y}))$, being $J$ the
Jacobian of the change, $d^3{\bf x}\equiv Jd^3{\bf y}$. For the
function $\tilde{\phi}(\bar{t},{\bf y})\equiv\sqrt{J}
{\phi}({\mathcal R}(\bar{t},{\bf y}))$,  the conjugated momentum is
$\tilde{{\xi}}(\bar{t},{\bf y})\equiv\f{\partial \tilde{\mathcal L}}
{\partial(\partial_{\bar{t}}\tilde{{\phi}})}=\sqrt{J}\partial_t\phi({\mathcal
R}(\bar{t},{\bf y}))$, and  the  Hamiltonian density
\begin{eqnarray}
\hspace*{-1mm} \tilde{{\mathcal H}}(\bar{t},{\bf y})=\f{1}{2}
\left(\tilde{{\xi}}^2 +J|\nabla_{\bf x}\phi|^2\right)
+\tilde{{\xi}}\left(\partial_{\bar{t}}\tilde{{\phi}}-\sqrt{J}\partial_t\phi\right).
\end{eqnarray}
In the coordinates $(t,{\bf x})$, after some calculations,
\begin{eqnarray}
&& \hspace*{-5mm} {{\mathcal H}}(t,{\bf x})= {{\mathcal E}}(t,{\bf
x}) +\xi(t,{\bf x})\langle \partial_s{\bf R}({\mathcal
R}^{-1}(t,{\bf x})) ,{\bf\nabla}_{\bf x}{{\phi}}(t,{\bf x})\rangle
\nonumber
\\ && \hspace*{10mm} +\f{1}{2}\left.\xi(t,{\bf x}) \phi(t,{\bf x})
\partial_s(\ln J)\right|_{{\mathcal R}^{-1}(t,{\bf x})}.
\end{eqnarray}
\indent For a single mirror which follows a prescribed trajectory
$(\epsilon g(t),t)$ in $1+1$ spacetime, we can set
$R(\bar{t},y)=y+\epsilon g(\bar{t})$, and obtain ${{\mathcal H}}(t,
x)= {{\mathcal E}}(t,x) +\epsilon
\dot{g}(t)\xi(t,x)\partial_x{{\phi}}(t, x)$.
\medskip

{\it Case of a single, partially transmitting mirror.}---We here
consider a single mirror in $1+1$ spacetime, following a prescribed
trajectory $(t,\epsilon g(t))$. When the mirror is at rest,
scattering is described by the matrix
\begin{eqnarray}S(\w)=\left(\begin{array}{cc}
{s}(\w)&{r}(\w)e^{-2i\w L}\\
{r}(\w)e^{2i\w L}&{s}(\w)\end{array}\right),
\label{matr1a}\end{eqnarray} where $x=L$ is the position of the
mirror. The $S$ matrix is taken to be real in the temporal domain,
causal, unitary, and the identity at high frequencies \cite{jr91}.
Specifically: (i) $S(-\w)=S^*(\w)$, (ii) $S(\w)$ is analytic for Im
$(\w)>0$, being $s(\w)$ and $r(\w)$ meromorphic (cut-off) functions
(the material's permitivity and resistivity), (iii)
$S(\w)S^{\dagger}(\w)=$ Id, and (iv) $S(\w)\rightarrow$ Id, when
$|\w|\rightarrow \infty$.

To reach the quantum theory from the Hamiltonian approach, we set
the mirror at $y=0$ in the above coordinates; the right and left
incident modes are
\begin{eqnarray}
&&\hspace*{-5mm}
\tilde{g}_{\w,R}(y)=\f{1}{\sqrt{4\pi\w}}\left\{{s}(\w) e^{-i\w y}
\theta(-y) \nonumber \right.\\ &&\hspace*{24mm} \left.
+\left[e^{-i\w y}+{r}(\w)e^{i \w y}\right] \theta(y)\right\}, \\ &&
\hspace*{-5mm} \tilde{g}_{\w,L}(y)=\f{1}{\sqrt{4\pi\w}}\left\{
\left[ e^{i\w
y}+{r}(\w)e^{-i \w y}\right]\theta(-y) \right. \nonumber \\
&&\hspace*{25mm}+ \left. {s}(\w)e^{i\w y} \theta(y)\right\}.
\end{eqnarray}
In the coordinates $(t,x)$ the instantaneous set of the right and
left incident eigenfunctions which generalize the set for a
perfectly reflecting mirror is $g_{\w,j}(t,x;\epsilon)\equiv
\tilde{g}_{\w,j}(x-\epsilon g(t)), \, j=R,L$. In general, we do
not know which is the part of the Hamiltonian that describes the
interaction between the field and the mirror. To get the
quantized theory, the energy of the field $E(t)=\int_{\R}dx{\mathcal
E}(t,x)$ which in presence of a single mirror does not depend on
$\epsilon$ must be considered as part of the free Hamiltonian of the
system. In the interaction picture, the Schr\"odinger
eq.~is
\begin{eqnarray}
i\partial_t|\Phi\rangle &=& \epsilon
\dot{g}(t)\int_{\R}dx\hat{\xi}_I(t,x;\epsilon)
\partial_x\hat{\phi}_I(t, x;\epsilon)
|\Phi\rangle \label{qe1} \\ &=& \epsilon
\dot{g}(t)\int_{\R}dx\hat{\xi}_I(t,x;0)
\partial_x\hat{\phi}_I(t, x;0)
|\Phi\rangle+{\mathcal O}(\epsilon^2), \nonumber
\end{eqnarray}
the average number of (quasi)particles \cite{gmm94} and the dynamical
energy (e.g., the energy of the created particles)
at time $t$ are, respectively,
\begin{eqnarray}
{\mathcal N}(t)&\equiv&\sum_{j=R,L}\int_0^{\infty}d\w \langle
0|\left({\mathcal T}^{t}\right)^{\dagger}
\hat{a}^{\dagger}_{\w,j}\hat{a}_{\w,j} {\mathcal T}^{t}|0\rangle,
\\
\langle\hat{E}(t)\rangle&\equiv&\sum_{j=R,L}\int_0^{\infty}d\w\w
\langle 0|\left({\mathcal T}^{t}\right)^{\dagger}
\hat{a}^{\dagger}_{\w,j}\hat{a}_{\w,j} {\mathcal T}^{t}|0\rangle,
\label{eet1}
\end{eqnarray}
being ${\mathcal T}^t$  the quantum evolution operator. A simple but
cumbersome calculation yields the following results
\begin{eqnarray}
{\mathcal N}(t)&=&\f{\epsilon^2} {2\pi^2}
\int_0^{\infty}\int_0^{\infty}\f{d\w d\w'\w \w'} { (\w+\w')^2}\left|
\widehat{\dot{g}\theta_t} (\w+\w')\right|^2 \label{N1} \label{nt1}\\
&&\hspace*{-5mm} \times
[|{r}(\w)+{r}^*(\w')|^2+|{s}(\w)-{s}^*(\w')|^2] +{\mathcal
O}(\epsilon^4),\nonumber  \\
\langle\hat{E}(t)\rangle&=&\f{\epsilon^2} {4\pi^2}
\int_0^{\infty}\int_0^{\infty}\f{d\w d\w'\w \w'} { (\w+\w')}\left|
\widehat{\dot{g}\theta_t}(\w+\w')\right|^2 \label{et1} \\ &&
\hspace*{-5mm}\times [|{r}(\w)+{r}^*(\w')|^2+|{s}(\w)-{s}^*(\w')|^2]
+{\mathcal O}(\epsilon^4),\nonumber
\end{eqnarray}
where $\theta_t$ is Heavyside's step function,
$\theta_t(\tau)=\theta(t-\tau)$, and $ \hat{f}$ the Fourier
transform of  $f$.

These two quantities are in general convergent. However, for the
seminal Davis-Fulling model \cite{fd76} of a single, perfectly
reflecting mirror, both quantities diverge when the mirror moves or
when its movement has discontinuities of some kind
\cite{mo70,sps98}. To obtain a finite energy, different
regularization techniques have been used. For instance, with a
frequency cut-off $e^{-\gamma\w}$, with $0<\gamma\ll 1$, the
regularized energy is $\langle\hat{E}(t;\gamma)\rangle=
\f{\epsilon^2}{6\pi}\left[
\f{\dot{g}^2(t)}{\pi\gamma}-\ddot{g}(t)\dot{g}(t)+
\int_0^t\ddot{g}^2(\tau)d\tau\right]$, and imposing that the kinetic
energy of the moving boundary be $\f{1}{2}
\left(M_{exp}-\f{1}{3\pi^2\gamma}\right)\epsilon^2\dot{g}^2(t)$,
with $M_{exp}$ the experimental mass of the mirror, some authors
conclude that the renormalized dynamical energy, namely
$\hat{E}_R(t)$, is \cite{fd76}-\cite{e93}
\begin{eqnarray}
\langle\hat{E}_R(t)\rangle \equiv
\f{\epsilon^2}{6\pi}\left[-\ddot{g}(t)\dot{g}(t)+
\int_0^t\ddot{g}^2(\tau)d\tau\right].\end{eqnarray} However when
$t\leq \delta$, with $0<\delta\ll 1$, this renormalized energy is
negative, which shows that, while the mirror moves, the renormalized
energy cannot be considered as the  energy of the produced particles
at time $t$ (cf. the paragraph after Eq.~(4.5) in \cite{fd76}). We
interpreted such results as implying that a perfectly reflecting
mirror is non-physical and decided to approach the problem by
considering instead a partially transmitting mirror, transparent to
high frequencies. Results are rewarding: in our Hamiltonian approach
Eqs.~(\ref{qe1}),~(\ref{eet1}),~(\ref{et1}), for the
radiation-reaction force, e.g.,
 the difference between the energy
density of the evolved vacuum state
on the left and right sides of the mirror,
 we
do get the right sign
\begin{eqnarray}
&&\hspace*{-5mm} \langle\hat{F}_{Ha}(t)\rangle =-\f{\epsilon}
{2\pi^2} \int_0^{\infty}\int_0^{\infty}\f{d\w d\w'\w \w'} {\w+\w'}
\Re\left[e^{-i(\w+\w')t} \right.  \label{fha1}
\\&&\hspace*{-5mm}\left.
\widehat{\dot{g}\theta_t}(\w+\w')\right]
[|{r}(\w)+{r}^*(\w')|^2+|{s}(\w)-{s}^*(\w')|^2] +{\mathcal
O}(\epsilon^2).\nonumber
\end{eqnarray}
Note this integral diverges for a perfect mirror ($r\equiv -1$,
$s\equiv 0$, ideal case), but nicely converges for our partially
transmitting (physical) one where $r(\w) \to 0$, $s(\w) \to 1$, as
$\w \to \infty$ (see (\ref{matr1a}) and ff). Energy conservation is
fulfilled: the dynamical energy at any time $t$ equals, with the
opposite sign, the work performed by the reaction force up to that
time $t$ \cite{fv82,be94}; in fact, from (\ref{et1}) and
(\ref{fha1}),
\begin{eqnarray}
\langle\hat{E}(t)\rangle=-\epsilon\int_0^t\langle\hat{F}_{Ha}(\tau)
\rangle\dot{g}(\tau)d\tau.\end{eqnarray}
 \indent {\it Comparison with
other results.}---First, we have repeated the calculations using the
Heisenberg picture approach of \cite{op01}. We have got the ``in''
modes when the mirror describes the prescribed trajectory
$(t,\epsilon g(t))$. Then, we have obtained the average number of
produced particles after the mirror returns to rest: \cite{bd82}
\begin{eqnarray}
\hspace*{-1mm} {\mathcal N}(t\geq T)=\sum_{i,j=R,L} \int_0^{\infty}
\int_0^{\infty}d\w d\w'\left|(\phi^{out }_{\w,i},\phi^{in
*}_{\w',j}) \right|^2\hspace*{-1mm},
\end{eqnarray}
by calculating the Bogoliubov coefficients $(\phi^{out
}_{\w,i},\phi^{in *}_{\w',j})$ in the null future infinity
${\mathcal I}^+$ (outgoing modes acquire a very simple expression in
${\mathcal I}^+$). The final result turns out to be exactly the same
expression (\ref{N1}).

The radiation-reaction force in the Heisenberg picture,
$\langle\hat{F}_H(t)\rangle$, is the difference between the energy
density of the ``in'' vacuum state on the left and right sides of the mirror. A simple
calculation shows that the energy density on both sides of the mirror
is \begin{eqnarray}&&\hspace{-3mm}
\langle\hat{\mathcal E}(t,x)\rangle =\int_0^{\infty}d\w \w \pm
\f{i\epsilon}{8\pi^2} \int_{\R^2}d\w d\w' \w \w' \hat{g}(\w+\w')
\chi(\w) \nonumber\\&&\hspace{-1mm} \times
[1+r(\w)r(\w')-s(\w)s(\w')]e^{-i(\w+\w')v} \theta(\pm(\epsilon
g(t)-x))\nonumber\\&& +{\mathcal O}(\epsilon^2),
\end{eqnarray}
$\chi(\w)\equiv\theta(\w)-\theta(-\w)$ being the sign function. Note
that the term of order $\epsilon$ is ill-defined, since the function
$\w \w' \hat{g}(\w+\w')[1+r(\w)r(\w')-s(\w)s(\w')]$ is not Lebesgue
integrable. Some regularization is needed to obtain a well-defined
quantity. Defining the regularized energy by
\begin{eqnarray}
&&\hspace*{-5mm}\langle\hat{\mathcal
E}(t,x;\gamma)\rangle\equiv\hspace*{-2mm}\sum_{j=R,L}\hspace*{-1mm}
\int_0^{\infty}\hspace*{-3mm} d\w e^{-\gamma\w}
\left[\partial_u\phi^{in}_{\w,j} (u,v;\gamma)\partial_u\phi^{in
*}_{\w,j}(u,v;\gamma)\right. \nonumber\\&&\hspace*{25mm}+\left.
\partial_v\phi^{in}_{\w,j}(u,v;\gamma)
\partial_v\phi^{in *}_{\w,j}(u,v;\gamma)\right],
\end{eqnarray}
 with the ``in'' modes regularized to obtain a
cut-off independent quantity \cite{he2}, in the Heisenberg picture
reads
\begin{eqnarray}&&\hspace*{-4mm}
\langle\hat{F}_H(t;\gamma)\rangle=\f{i\epsilon}{8\pi^2}
\int_{\R^2}d\w d\w' \w \w' \hat{g}(\w+\w') [\chi(\w)+\chi(\w')]
\nonumber\\&&\times[1+r(\w)r(\w')-s(\w)s(\w')]e^{-\gamma(|\w|+|\w'|)}
e^{-i(\w+\w')t}\nonumber\\&& \hspace*{3mm}+{\mathcal O}(\epsilon^2).
\end{eqnarray}
This converges and is cut-off independent, and a possible definition
of the  renormalized radiation-reaction force is
\begin{eqnarray}&&\hspace*{-3mm}
\langle\hat{F}_{H,ren}(t)\rangle=\f{i\epsilon}{8\pi^2}
\int_{\R^2}d\w d\w' \w \w' \hat{g}(\w+\w') [\chi(\w)+\chi(\w')]
\nonumber\\&&\hspace*{1mm}\times[1+r(\w)r(\w')-s(\w)s(\w')]
e^{-i(\w+\w')t}+{\mathcal O}(\epsilon^2). \label{fhren1}
\end{eqnarray}
In general, this formula disagrees with the radiation-reaction force
(\ref{fha1}) which was obtained using the Hamiltonian approach.
Moreover, we have been able to prove (details will be provided
elsewhere \cite{he2}) that the force (\ref{fhren1}) coincides with
the radiation-reaction force calculated by Jaekel and Reynaud
\cite{jr92} after renormalization:
$\langle\hat{F}_{J,R,ren}(t)\rangle\equiv
\langle\hat{F}_{H,ren}(t)\rangle$. We thus conclude that the method
of Jaekel and Reynaud is equivalent  to the quantum theory in the
Heisenberg picture. Furthermore, note that
$\epsilon\int_{\R}dt\langle\hat{F}_{Ha}(t)\rangle \dot{g}(t)=
\epsilon\int_{\R}dt\langle\hat{F}_{J,R,ren}(t)\rangle \dot{g}(t)$,
and this identity proves that the dissipative parts of
$\langle\hat{F}_{Ha}(t)\rangle$ and
$\langle\hat{F}_{J,R,ren}(t)\rangle$  always agree.

However, in several situations the reactive parts do not match. For
instance if ${r}(w)=-\f{i\alpha}{\w+i\alpha}$ and
${s}(w)=\f{\w}{\w+i\alpha}$ with $\alpha>0$, there is the relation
\begin{eqnarray}
\langle\hat{F}_{Ha}(t)\rangle=-\f{\alpha\epsilon}{2\pi}\ddot{g}(t) +
\langle\hat{F}_{J,R,ren}(t)\rangle,
\end{eqnarray}
where
\begin{eqnarray}
\langle\hat{F}_{J,R,ren}(t)\rangle=\f{\alpha\epsilon}{\pi}\hspace*{-1mm}
\int_1^{\infty}\hspace*{-4mm}dz\hspace*{-1mm}\int_{-\infty}^t\hspace*{-5mm}
d\tau(z^{-2}-z^{-3}) e^{-\alpha z(t-\tau)}\dddot{g}(\tau).\ \
\label{jrr1}
\end{eqnarray}
The two forces differ in a reactive term. Now the crucial point is
that, during the movement of the mirror, the work done by the motion
force $\langle\hat{F}_{J,R,ren}(t)\rangle$ is {\it not} a negative
quantity. Consequently, the dynamical energy is not  positive and a
meaningless result is obtained because the dynamical energy is the
energy of the produced particle. To avoid this difficulty, the
reactive term $-\f{\alpha\epsilon}{2\pi}\ddot{g}(t)$, which most
naturally appears in the Hamiltonian formulation for a partially
transmitting mirror, comes to rescue and renders a physically
meaningful result.

Barton and Calogeracos  \cite{bc95} (see also \cite{dmn1}) studied the case
${r}(w)=-\f{i\alpha}{\w+i\alpha}$, ${s}(w)=\f{\w}{\w+i\alpha}$, with
$\alpha>0$. The interaction between  field and  mirror can be
described there by the Lagrangian density $
\f{1}{2}\left[(\partial_t\phi)^2-(\partial_x\phi)^2\right]
-\alpha\phi^2 \delta(x-\epsilon g(t)). $ With the Hamiltonian method
we have obtained the corresponding quantized Hamiltonian, and
\begin{eqnarray}
\hspace*{-2mm}\int_{\R}\hspace*{-2mm}dx\hat{{\mathcal E}}(t,x)
+\alpha\hat{\phi}^2(t,\epsilon g(t))=
\hspace*{-3mm}\sum_{j=L,R}\hspace*{-1mm}\int_0^{\infty}\hspace*{-4mm}d\w
\w (\hat{a}^{\dagger}_{\w,j} \hat{a}_{\w,j}+1/2), \ \ \label{intr1}
\end{eqnarray}
from where we conclude that the quantum equation, in the interaction
picture, is given by (\ref{qe1}). This leads, for these reflection and
transmission coefficients, back to our formulae (\ref{nt1}),
(\ref{et1}), (\ref{fha1}). However, two important differences exist
between those and our results. First,  to obtain the Schr\"odinger
equation, these authors make a unitary transformation which does not
seem easily generalizable to the case of two moving mirrors. And
second, in \cite{bc95}, following \cite{e93,be94},  a mass
renormalization is performed ---in order to eliminate the reactive
part of the motion force--- where the energy of the field is not a
positive quantity at any time $t$. Again, the concept of particle
is ill-defined during the mirror's displacement.
\medskip

{\it Two partially transmitting mirrors.}---We have finally extended
our method to the case of two moving mirrors that follow prescribed
trajectories, $(t,L_j(t;\epsilon))$, where $L_j(t;\epsilon)\equiv
L_j+\epsilon g_j(t)$, with $j=1,2$, assuming that
$L_1(t;\epsilon)<L_2(t;\epsilon)$, for all $t\in \R$. In this case
it is impossible, in practice, to work in the Heisenberg picture,
because it is extremely difficult to obtain the ``in'' and ``out''
mode functions in the presence of the two moving mirrors. Instead,
in order to get the dissipative part of the motion force, the number
of radiated particles, and their energy, the approach of Jaekel and
Reynaud can be used, which starts from the effective Hamiltonian
$\hat{H}_{J,R}\equiv -\sum_{j=1,2}\epsilon g_j(t)\hat{F}_j(t)$,
where $\hat{F}_j(t)\equiv \lim_{\delta\rightarrow 0}
\left[\hat{\mathcal E}(t,L_j-|\delta|)- \hat{\mathcal
E}(t,L_j+|\delta|)\right]$ is the force operator at the point
$x=L_j$ \cite{jr9236}. However this method does not seem useful to
obtain the reactive part of the motion force and the dynamical
energy while the mirrors move. As before, in order to get those
quantities we are led to use our Hamiltonian approach. This demands
now considerable effort \cite{he2}, e.g., generalizing the model \cite{bc95},
described by the Lagrangian density
$
\f{1}{2}\left[(\partial_t\phi)^2-(\partial_x\phi)^2\right]
-\sum_{j=1,2}\alpha_j\phi^2 \delta(x-L_j(t;\epsilon))$, the
Hamiltonian in the interaction picture has the form
\begin{eqnarray}&&\hspace*{-5.5mm}
\hat{H}_I(t)=-
\f{\epsilon[g_2(t)-g_1(t)]}{L_2-L_1}\hspace*{-1mm}\left[\hspace*{-.5mm}\int_{\R}\hspace*{-2mm}dy
(\partial_y\widehat{\tilde{\phi}}_I(y))^2
\hspace*{-1mm}+\hspace*{-2mm}\sum_{j=1,2}\hspace*{-2mm}\alpha_j
(\widehat{\tilde{\phi}}_I(L_j))^2\right]
\nonumber\\&&\hspace*{-2mm}+ \epsilon\hspace*{-1mm}\sum_{j=1,2}
\int_{\R}\hspace*{-1mm}dy\f{(-1)^j\dot{g}_j(t)\widehat{\tilde{\xi}}_I(y)}{
L_2-L_1}\hspace*{-1mm}\left[
\partial_y\widehat{\tilde{\phi}}_I(y)(y-\bar{L}_{j})
+\f{1}{2}\widehat{\tilde{\phi}}_I(y)\right]\nonumber\\&&\hspace*{12mm}+
{\mathcal O}(\epsilon^2),
\end{eqnarray}
in terms of free quantum fields defined from an expansion in terms
of left and right incident eigenfunctions.  Our dissipative part of
the motion force \cite{he2} coincides with the one obtained in
\cite{jr9236}. For times $\tau$ larger than the stopping time, our
quantum evolution operator is ${\mathcal
T}^{\tau}=\mbox{Id}-i\int_{\R}dt\hat{H}_I(t)$. Using results from
\cite{bc95}, we obtain explicitly that, for times $\tau$ larger than
the stopping time,
$
{\mathcal T}^{\tau}=\mbox{Id}+i\epsilon\sum_{j=1,2}\int_{\R}dt
g_j(t)\hat{F}_j(t),
$
as it should be. We see no basic obstruction to extend our procedure
to higher dimensions and fields of any kind.

We should mention, to finish, that there are proposals to detect
the radiated photons, although the reactive part and the possible
deviations from conservative motion seem out of experimental reach yet
\cite{kbo1}.


We thank Gabriel Barton, Steve Fulling and Roberto Onofrio for
helpful discussions and correspondence, and the last also for
hospitality. This work was supported in part by MEC (Spain),
projects MTM2005-07660-C02-01, BFM2003-00620, PR2006-0145, and by
AGAUR (Gene\-ra\-litat de Catalunya), contract 2005SGR-00790.

\end{document}